# Tie-RBAC: An application of RBAC to Social Networks

*Antonio Tapiador, Diego Carrera, Joaquín Salvachúa*

*Universidad Politécnica de Madrid*

**Abstract**

This paper explores the application of role-based access control to social networks, from the perspective of social network analysis. Each tie, composed of a relation, a sender and a receiver, involves the sender's assignation of the receiver to a role with permissions. The model is not constrained to system-defined relations and lets users define them unilaterally. It benefits of RBAC's advantages, such as policy neutrality, simplification of security administration and permissions on other roles. Tie-RBAC has been implemented in a core for building social network sites, Social Stream.

# 1. Introduction

New social networking paradigms[1] show people and organizations sharing resources with different - usually independent - groups, each one containing different levels of weak and strong ties between their members. But current social network management systems (SNMS) apply a simple security model based on friends, friends of friends, like Facebook or followers of Twitter, etc. The word *friend* is too coarse to embrace the richness of social relationships. Social entities are individuals, but also groups, organizations, institutions and even social events. They share resources at different levels, based on their own defined strong and weak ties. Each person describes her relationships using different words, such as *colleague*, *classmate*, *business partner* or *acquaintance*. It is the same case for organizations, which define positions such as *administrator*, *operator*, *organizer*, or *participant*. All social entities have ties between them. Ties are also established between social entities of different types.

Relationships are often not reciprocal. A *computer science department* nominates *Alice* as *assistant professor*, but the opposite cannot be possible. *Alice's friend*, *Bob*, may consider her as a *partner*. The access control model should support relation names and properties being defined and customized by users. These customizations should be unidirectional and flexible enough to not constrain access control design. Besides, relation types are associated to different levels of information disclosure. People usually share more private information with *close friends* than with other type of contacts. The *professors* of a *department* have more privileges on it than its *students*. SNMS should support assigning permissions at the same time the contact is created and classified.

This paper introduces an access control model supporting these features. Tie-RBAC arises from the application of the well-known and successful RBAC model to the comprehensive discipline of social network analysis. It takes advantage of well-known advantages of RBAC: policy-neutrality, simplification of security administration and permissions on roles. Besides, it allows users to grant permissions to other actors at the same time they create contacts in the SNMS.

After a description of the model, the paper shows its application on a real case of video-conferences based on a core for building social network websites, to finish with some conclusions.

# 2. Methodology

Tie-RBAC is the application of role based access control (RBAC) to social network analysis (SNA). The first is a modern and successful access control model that provides an efficient way to manage permissions, while the second provides a comprehensive body of concepts and methods for modeling social networks. To the best of our knowledge, this is the first time RBAC is applied to social networks.

$$RBAC + SNA = Tie\text{-}RBAC$$

## 2.1 Social Network Analysis

Social network analysis (SNA) provides the foundations for understanding the linkages between social entities and the implications of these linkages [12]. It provides researchers with a full set of methods for the analysis of social networks, as well as a collection of solid concepts.

In SNA, social entities are referred to as actors. An *actor* is a discrete **individual**, but it can be also **a group**, **a department**, **an organization**, **even a nation or state in the world**. All of them are considered entities in social literature. Actors are linked by social ties. A *tie* is made up of **two actors and the type of the tie**. Actors are ordered, the first of them is the *sender* of the tie and the second one is the *receiver* of the tie. A social network with ties between only one type of actors is called a one-mode network. These are the only ones in current SNMS implementations and access control models which consider only ties between users and do not include other types of social entities.

The most common types of ties between actors include affective (friendship, liking, respect), formal or biological relationships (authority, kinship), transfer of material resources (transactions, lending and borrowing), messages or conversations, physical connection and affiliation to same organizations. A *relation* is the set of all **ties of the same type** between actors in the network. Examples are the set of friendships in a classroom or the set of commercial transactions between nations in the world. Two actors can have or not ties in different relations. For example, two children in the same class may be friends but not be seated together at the same desk.

Relations can be reciprocal. If there is one tie with a reciprocal relation, there must be another tie with the actors in reverse order. This is the case of a *friend* relation managed by Facebook, where both parts must accept the tie before it is established. If there is a tie of friendship between *Alice* and *Bob*, there will be other between *Bob* and *Alice*. Other relations are not reciprocal, the reciprocal tie may or may not exist. This is the case of the *follower* relation on Twitter. *Bob* is following *Charlie*, but *Charlie* may or may not be following *Bob*. **Tie-RBAC is based on non-reciprocal ties**. The system must provide actors the flexibility to define their own relations. We do not rely on what *Alice* defines as *friend* as the same thing as *Bob*'s notion of *friend*. Besides, there may be different points of view on the same relationship between two people. *Alice* may consider Bob a *friend*, while *Bob* considers Alice a *partner*. We find here a significant difference between SNA and SNMS design. The SNA methods analyze a social network in order to obtain the structure of the network. Data is recollected and treated to understand how the social network works. SNMS design goes the other way. It establishes the rules along with the generation of data. It sets up relations before actors inter-act. For this reason, we propose a flexible model where letting actors find out their most suitable relations is important.

## 2.2 Role Based Access Control

Role-Based Access Control (RBAC) [4] is a conceptual model that helps to manage access control information using the metaphor of roles. In RBAC, users are assigned to roles. Permissions - actions performed on an object - are also assigned to roles. Thus, users acquire permissions by being members of roles. This simplifies the management of privacy policies, by assigning them to roles instead of single entities. The RBAC model is organized into four components, each providing more features to the model. Core RBAC captures the basic features of users, roles and permissions. Hierarchical RBAC, the second component, introduces role hierarchies that reflect the organization's lines of authority and responsibility. Privileges are inherited through the hierarchy. When a user is assigned to a role at the top of the hierarchy, she will also have permissions belonging to roles further down in the hierarchy. The other two components are the static separation of duty and dynamic separation of duty.

RBAC provides several well-recognized advantages over other access control models, such as Discretionary Access Control (DAC) and Mandatory Access Control (MAC) [8]:

- Arbitrary, organization-specific security policies. RBAC is "policy neutral". A wide range of security policies can by defined, including DAC, MAC and user-specific. This point fits our requirements on user-centric approach.
- Simplification of security administration. It supports moving users to new roles instead of revoking individual permissions.
- Special administrative roles can be designed to manage other roles. This aspect provides a powerful feature never seen before in SNMS. The capability of assigning representatives in a social network. An example are important and busy people which can delegate some of their administrative tasks on other people in the SNMS.

## 2.3 RBAC application to social networks

The key point of our contribution is the application of RBAC to social networks. In Tie-RBAC, the tie element defined in SNA is re-interpreted in the context of the RBAC model as the assignation of a user to a role. Actors define their custom relations (*friend*, *partner* etc.), which are equivalent to roles. Actors assign permissions (*read wall*, *post to wall*, *represent*, etc.) to relations. Additionally, actors make ties using those relations. **Each tie is equivalent to the assignation of an actor to a role-relation (figure 1).** The sender of the tie is the entity which grants the privileges on their objects when establishing the tie. The receiver of the tie is the entity assigned to the role, which gains privileges on the sender's objects. The relation of the tie, that is defined by the sender, is the role. Note that both the sender and receiver are actors, so they can be users, but also group, organizations or any other type of social entity.

In the example, *Alice* defines the relation *friend*. She grants *read wall* and *post to wall* permissions to *friend*. When she establishes the tie to *Bob*, she chooses *friend* as the relation of the tie. At the same time, she is granting *Bob* the permissions of reading and posting to her wall. On the other hand, the *computer science department* defines a relation *delegate*, and assigns the permission *represent* to it. Then, it adds *Charlie* as a *delegate*, so he can now represent the department in the application.

The last example has the chicken and egg problem. If there are not delegates, who will create the delegate relation and will establish the ties? One solution are default relations, defined by the administrator of the application. Tie-RBAC model is general enough to support user-centric definition of relations. But administrator-defined relations can be supported in the system as well.

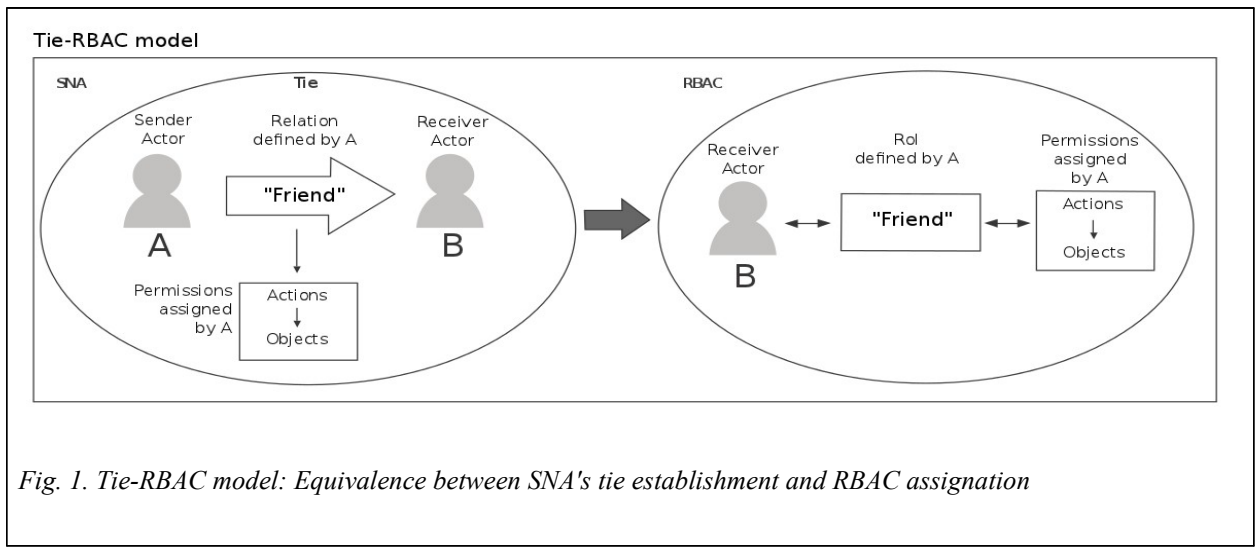

Fig. 1. Tie-RBAC model: Equivalence between SNA's tie establishment and RBAC assignation

## 3. Results

We have implemented Tie-PRBAC in Social Stream[4], a Ruby on Rails engine for building social network websites. Ruby on Rails, the framework for agile web development, provides support for engines, a powerful kind of plug-in that supports mounting web applications. Social Stream is a base application providing social networking functionalities in order to create social websites. Our goal with Social Stream is to provide social actors and web developers with a tool to build websites with social network features. These websites may be contact-oriented or content-oriented, specialized in a field such as travelling, source code sharing, etc. In our case, we are using Social Stream in the re-factorization of a real world application, GlobalPlaza[2], a website oriented to the organization of online video-conferences. It is developed in the context of the Global Project[3], a research project supported by the European Commission's seventh framework program. We found strategic adding social networking features in order to improve engagement, enhance user awareness and stimulate communities around the site.

In GlobalPlaza, there are two types of social entities besides users, spaces and events. A space is a wrapper for research groups, organizations and any kind of institution. Events are also considered social entities, as they have their own public image, provide a place for user interactions and have ties with other entities, both users and spaces. System-defined default relations for users include *friend*, *acquaintance* and *public*. Default space relations include *administrator*, *member*, *follower*, *partner* and *public*. Default event relations include *organizer*, *participant* and *audience*. The application of the model to this real world case proves the flexibility of our model. There are two types of actors beyond users: spaces and events. Different kinds of relations are defined, depending on actor type. Permissions are defined for those relations, improving flexibility in permissions management. Besides, each actor is able to define their own kind of relations and customize them with their own permissions.

## 4. Related work

Current practices in SNMS access control are reviewed in literature [2] [3] [7] [9] [10] [11]. Present approaches adopt a Discretionary Access Control (DAC) point of view. Resources have an attribute used to control its access. This attribute usually takes one of the following values: private, friends, friends-of-friends and public. Although these are the most common values, SNMSs actually manage more types, e.g. follower, colleague, classmate and business partner. Carminati et al. [3] provide a full review of them. This popular method is attributed to the straightforwardness and ease of its implementation, and the best balance between ease-of-use and flexibility [2][3][7]. But it introduces the following limitations: it is too coarse and using the friends' relation for access control forces users to choose between protecting their privacy and appearing popular.

Some recent proposals try to overcome these limitations. Squicciarini et al. carries out extensive work in the field. Their contributions include a solution with automated ways of sharing images based on an extended notion of content ownership, using a game theory based mechanism [10], "web-traveler policies" attached to images in order to restrict access to them in SNMS [11] and PriMa, a privacy protection mechanism which supports semi-automated generation for access rules for user profile information [9]. All the work is based on policy specifications on resources, which take a Discretionary Access Control (DAC) approach.

Ali et al. and Carminati et al. base their solutions on trust. Ali et al.'s approach [1] define an evaluation of user reputation based on trust relationships with other users. Resources are given a confidence level. They will be accessed by users with more or equal reputation level. The Carminati et al. solution [3] is more sophisticated. They propose a complete semi-decentralized solution based on three parameters, i.e. type of link, path depth and trust value of the link. The model supports decentralized SNMS with a centralized certificate authority that asserts the validity of the links. Both of them rely on trust, measured as a numeric value.

A different approach is taken by Fong et al. [5]. They describe a full algebraic model for access control, claiming Facebook to be just an instance of it. Their model is abstract to large degree. Authorization is based on two issues. The first is the communication history, the set of events that happen between each user, e.g., invitations to participate in the social network. The second one is acquaintance topology, the set of relationships between users stored in the persistent layer of the SNMS. They introduce the latter to simplify the former. As the authors recognize, consuming all the communication history in order to evaluate an authorization request is intractable. However, significant communication events could also be saved as relationships, which would simplify the model considerably more. Relation-based access control is also suggested by Gates as a requirement for access control [6].

We have found that current work lacks several interesting features. There is no way to act from inside other social actors different from users. Groups and other actors are an important part of social behavior, but they are not the subjects of current access control models. In some solutions, groups are considered as part of the DAC policy definitions but no solution considers them as first-class actors. Instead of assigning a numeric value of trust to links, our approach focuses on the type of relationship as the channel for the transfer of information. Relations are ordered by their strength, assigned permissions to them in the same way as RBAC does. Finally, none of the models includes the advantages of RBAC [8].

## 5. Conclusions and future work

Tie-RBAC brings new possibilities to the field of access control in social networks. Based on RBAC advantages, it supports users representing other types of social entities, such as groups or organizations. It supports a variety of policy specifications. It provides actors with the ability to define their own relations, adapted to their own slang or field of activity. These relations are independent of system and administrator defaults. RBAC simplifies security administration. Usability tests with real users is planned as future work.

We were able to define them in the field of a video-conference system, including relationships between other social entities beyond individuals, such as social events. This model provides a powerful method for actors in the social network to concede access rights to their contacts at the same time as they establish relationships.

Finally, our model is a centered web social network (WBSN). Carminati and others raise several trust issues regarding a centralized WBSN. In a centered WBSN, the access control is also centralized. The SNMS stores access control policies, and it is responsible for enforcing access control. Release data to unauthorized users could be possible [3]. Users must trust the SNMS administrates their data correctly. This is a limitation of Tie-PRBAC. We also find a decentralization of our model as an interesting future work.